\documentclass[preprint,notoc,float]{JHEP3}
\usepackage{epsfig,axodraw,amssymb,amsmath,citesort}

\input bbh.mac
\input bbh.fig
\input bbh.pre

\begin{document}

\tableofcontents
\newpage


\section{Introduction}

The search for the Higgs boson(s) has been one of the central problems in
high energy physics. It is not only important to find the Higgs boson(s), it
is just as important to understand its properties, including couplings. These
will reveal the underlying electroweak symmetry breaking mechanism, be it
Standard Model (SM)
or beyond the SM. In order to distinguish between different physical
mechanisms, it is essential to assess the inherent uncertainties of
theoretical predictions based on various underlying physics scenarios.

In this regard, one of the first physical quantities that we need to gain a
good theoretical control of is the production cross section of the Higgs
particles. A lot of work has been done on higher order QCD corrections to the
Higgs boson production cross section in SM and in Supersymmetry models. These
have considerably reduced the \textquotedblleft theoretical
uncertainties\textquotedblright\ of the calculated cross sections, usually
estimated by varying the renormalization and factorization scales over some
range (say, by a factor of 2). We shall refer to this type of uncertainties,
in short, as \emph{scale uncertainties}. \ However, the total uncertainty of
the predicted Higgs production cross sections also includes uncertainties due
to parton distribution functions~(PDFs)---in short,  \emph{PDF
uncertainties}. These can be significant, or even dominant, compared to the
scale uncertainties, depending on the theoretical model and the model
parameters. \

In this paper we perform a detailed study of the PDF uncertainty for
inclusive Higgs boson production at the
Fermilab Tevatron and CERN Large Hadron Collider (LHC) and compare them
with existing estimates of the scale uncertainties. Particular attention is
given to the $b$-quark initiated Higgs boson production mechanism. Whereas
the $b\bar{b}\rightarrow \mH$ process is relatively insignificant compared
to $gg\toploop\mH$ (via a top quark loop)\ in the SM,\footnote{%
In models with more than one neutral Higgs boson, the symbol $\mH$ is used
in the generic sense to represent any one of the neutral Higgses, such as \{$%
h,H,A$\} in MSSM, cf.~Sec.~2.} this is not so in many models beyond the SM,
such as the Minimal Supersymmetric Standard Model (MSSM). The presence of
several vacuum expectation values in these models can lead to a large
enhancement of the Yukawa coupling of the $b$-quark to some of the Higgs
bosons. These scenarios offer attractive opportunities for Higgs boson
searches at hadron colliders. It is therefore important to have reliable
estimates of all uncertainties in the calculation of the $b\bar{b}\rightarrow
\mH$ cross section.

In Section \ref{sec:SubProc}, we describe general features of the models with enhanced $b%
\bar{b}\mH$ coupling, as well as the specific model that is used in our
numerical study---the production of the CP-odd Higgs boson $A$ in MSSM. In
Section \ref{sec:ProdMech}, we discuss the relevant (scheme-dependent) QCD subprocess for $b%
\bar{b}\mH$-coupling-enhanced Higgs production in these models; and the
particular scheme used in our PDF uncertainty study. In Section
\ref{sec:Uncert}, we present the results on these uncertainties for the
inclusive Higgs boson production at Run II of the Tevatron and the LHC, and
compare them to the scale uncertainties available in the literature. We find
that the PDF uncertainties dominate over the scale uncertainties, except for
low Higgs masses at LHC. The PDF uncertainties are calculated by the robust
Lagrange multiplier method within the CTEQ global analysis framework, and
compared to those obtained by the (more approximate) Hessian method used
before. We show that the results of these two methods are consistent with
each other. Section \ref{sec:Concl} states our conclusions.



\section{Higgs Production in the SM and in MSSM}
\label{sec:ProdXsec}

\subsection{$b\bar{b}\protect\mH$ Yukawa Coupling Enhanced Higgs Production}
\label{sec:SubProc}

\figProdMech

 In the Standard Model, Higgs production by the partonic process \bbH is
small compared to production by \ggH, cf. Fig.~\ref{ProdMech}, because the $b%
\bar{b}H$ coupling is small compared to the $t\bar{t}H$ coupling, and
because the $b$-quark distribution is much smaller than the gluon
distribution. However, the two production mechanisms can become comparable
in beyond the Standard Model theories in which the $b\bar{b}\mH$ Yukawa
coupling is substantially enhanced with respect to $t\bar{t}\mH$~\cite%
{Barnett:1984zy,Pandita:1984hf,Hung:1987yj}. For example, in
the MSSM, there are two Higgs doublet superfields, with two independent
vacuum expectation values (VEVs): $v_{u}$ and $v_{d}$. While the sum of the
squares of these VEVs is fixed by the well-known $Z$ boson mass, their
ratio, denoted by $\tan \beta =v_{u}/v_{d}$, is a free parameter of the
model. After spontaneous symmetry breaking, there are 5 physical particles
in the Higgs sector: $h$ (light), $H$ (heavy), $A$ (pseudoscalar) and $%
H^{\pm }$(charged). An important feature of this model is that
with high values of $\tan \beta$, the Yukawa
couplings of the $b$-quark to the neutral Higgses are enhanced by a factor $%
1/\cos \beta $ compared to their SM value. The $t$- and $b$- quark Yukawa
couplings can be written as
\begin{eqnarray}
h_{t} &=&{\frac{\sqrt{2}\,m_{t}}{v_{u}}}={\frac{\sqrt{2}\,m_{t}}{v\sin \beta
}}, \\
h_{b} &=&{\frac{\sqrt{2}\,m_{b}}{v_{d}}}={\frac{\sqrt{2}\,m_{b}}{v\cos \beta
}},  \label{htdef}
\end{eqnarray}%
and the MSSM Yukawa coupling of the Higgs bosons to $t$- and $b$- quarks, $%
Y_{\mH f\bar{f}}$, relative to the SM one $Y_{hf\bar{f}%
}^{SM}=gm_{f}/2m_{W}=m_{f}/v$, takes the form
\begin{eqnarray}
Y_{ht\bar{t}}=({\ \cos \alpha /\sin \beta }) &&Y_{ht\bar{t}}^{SM} \nonumber
 \\
Y_{Ht\bar{t}}=({\ \sin \alpha /\sin \beta }) &&Y_{ht\bar{t}}^{SM}  \nonumber \\
Y_{At\bar{t}}={\ \cot \beta } \ \gamma _{5} &&Y_{ht\bar{t}}^{SM}  \label{eq-yukawa1} \\
Y_{hb\bar{b}}={-(\sin \alpha /\cos \beta )} &&Y_{hb\bar{b}}^{SM}  \nonumber \\
Y_{Hb\bar{b}}=({\ \cos \alpha /\cos \beta }) &&Y_{hb\bar{b}}^{SM}  \nonumber \\
Y_{Ab\bar{b}}={\ \tan \beta } \ \gamma _{5} &&Y_{ht\bar{t}}^{SM}   \label{eq-yukawa2}
\end{eqnarray}%
where $\alpha $ is the mixing angle of two CP-even Higgs bosons, and the
weak scale $v\ $equal to $246$~GeV.

In the MSSM, the Higgs mass is a function of $m_{A}$ and $%
\tan \beta $. The relatively high lower limit on the Higgs boson mass
deduced from LEP data~\cite{unknown:2001xx} favors scenarios with high $\tan
\beta$. Also, theoretically, high $\tan \beta $ scenarios are highly
motivated by SO(10) SUSY GUTs (see e.g. Refs.~\cite%
{Mohapatra:1999vv,Gell-Mann:1976pg,Fritzsch:1974nn,Raby:2004br,Altarelli:2004za,Ananthanarayan:1991xp,Anderson:1993fe,Carena:1994bv,Rattazzi:1995gk,Ananthanarayan:1994qt,Blazek:1999ue,Blazek:2001sb,Blazek:2002ta,Baer:1999mc,Baer:2000jj,Baer:2001yy,Auto:2003ys,Auto:2004km}%
). Thus, it is important to explore the phenomenological consequences of
Higgs production by enhanced $b\bar{b}\mH$ Yukawa coupling mechanisms at the
Tevatron and the LHC.

To make our study less dependent on SUSY parameters, we focus on production of
the CP-odd Higgs particle $A$. As seen from Eq.~(\ref{eq-yukawa1} \& \ref{eq-yukawa2}), the Yukawa
couplings of $A$ to the heavy quarks are independent of the Higgs mixing angle
$\alpha $; and the $Ab\bar{b}$ ($At\bar{t}$) coupling is enhanced (suppressed)
by the factor $\tan \beta $. For simplicity, we shall not consider SUSY-QCD
and SUSY-EW corrections to the Yukawa couplings, which
could be significant in specific regions of SUSY parameter space.\footnote{%
One should notice that squark contributions to \ggmH could be important~\cite%
{Dawson:1996xz,Harlander:2003bb,Harlander:2003kf,Harlander:2004tp}. Even the %
\ggA process under study (which does not receive squark contribution at
leading order if CP is conserved) can have sizable SUSY-QCD and SUSY-EW
corrections to the Yukawa couplings, which could enhance both \ggA and \bbA %
processes (see e.g.\ Refs.~\cite{Carena:2000uj,Belyaev:2002eq}.)}



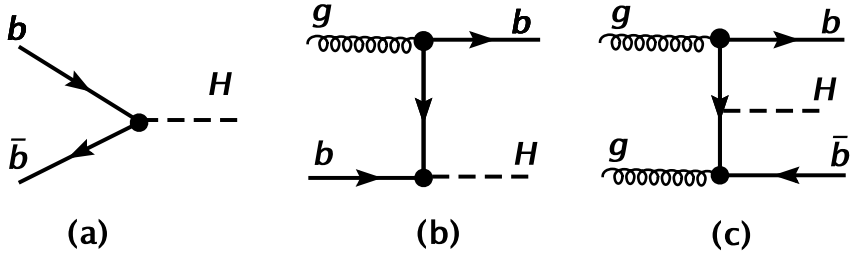
\subsection{Production Mechanisms with Enhanced $b\bar{b}\mH$ Coupling}

\label{sec:ProdMech}

The simplest partonic processes contributing to inclusive Higgs production
with enhanced $b\bar{b}\mH$ coupling are represented by the tree diagrams of
Fig.~(\ref{QcdTree}): (a) $b\bar{b}\rightarrow \mH$; (b) $gb\rightarrow \mH %
b $; and (c) $gg\rightarrow b\bar{b}\mH$. In QCD, these are not independent
production mechanisms, since $b$-partons inside the hadron beam/target arise
from QCD evolution (splitting) of gluons, and gluons radiate off quarks \cite%
{Collins:1986mp,Olness:1988ub,Barnett:1987jw}. The three processes (a,b,c)
all give rise to the \emph{same hadronic final states}, with two B-mesons
appearing in different, but overlapping, regions of phase space---either
as beam/target remnants or as high $p_{T}$ particles. The distinction
between the three processes depends very much on the factorization scheme
adopted for the QCD calculation.\footnote{%
For a detailed explanation and complete references, see the appendix of
\cite{Amundson:2000vg}. \label{fn:schemes}}

\figProdQcdTree

For example, in the (fixed) 4-flavor scheme which is often used in $b$-quark
production calculations, there are no $b$-partons \emph{by definition};
hence (a) and (b) are absent, so the gluon-fusion process (c) is the only
one contributing. By comparison, in the 5-flavor scheme, all three
subprocesses contribute; and, in fact, all three are numerically comparable
in spite of the differences in the nominal number of powers of $\alpha _{s}$
present ($0,1,2$ respectively).\footnote{%
This conclusion holds,
except at asymptotic energies, beyond that of foreseeable
accelerators, when the heavy quark parton
distributions become comparable to that
of the light quarks.} This is because the magnitude of the gluon
distribution $g(x,Q)$ is much larger than the heavy quark distribution $%
b(x,Q)$---by at least a factor of $\alpha _{s}^{-1}$ at currently
available energy scales---hence compensating for the $\alpha _{s}$
factors. In this scheme, when all three subprocesses are included, the
overlap between them needs to be properly taken into account. This is most
conveniently done by taking the $m_{b}\rightarrow 0$ limit in the QCD
calculations, while keeping $m_{b}\neq 0$ only for Yukawa couplings.

These {$\mH$}${b\bar{b}}$ processes have been extensively studied in recent
years for MSSM and for other beyond SM scenarios with similar enhanced $b%
\bar{b}\mH$ couplings~\cite%
{Diaz-Cruz:1998qc,Balazs:1998nt,Dicus:1998hs,Campbell:2002zm,Maltoni:2003pn,Boos:2003yi,Hou:2003fm,Dittmaier:2003ej,Harlander:2003ai,Dawson:2003kb,Kramer:2004ie,Field:2004nc,Dawson:2004sh}%
. Calculations in the 5-flavor scheme have been carried out to the 2-loop
level \cite{Harlander:2003ai}, which considerably reduced the theoretical
uncertainty due to the perturbative expansion, as estimated by the residual
scale dependence. \ Comparison of results obtained in the 4- and 5-flavor
schemes has also been carried out \cite{Dawson:2004sh}. It shows consistency
between the two schemes in the energy region of the Tevatron and the LHC.

For the purpose of this paper---assessing the range of uncertainties
associated with input parton distribution functions---it suffices
to\figNloBbh
calculate the production cross section of the Higgs particle $A$, $p\overset{%
(-)}{p}\rightarrow A+X$, to the one-loop level in QCD in the 5-flavor scheme,
which will be referred as next-to-leading order (NLO) cross section in this
work. The Feynman diagrams representing the partonic subprocesses included in
the calculation are shown in Fig.~\ref{fig:NloBbh}. The numerical calculation
has been carried out with the program developed in \cite{Balazs:1998sb},
where the QCD-improved (running) Yukawa couplings have been used.

\subsection{Cross Sections for Higgs Production in SM and MSSM}

To make the discussion concrete, we now compare the order of magnitudes of
Higgs boson production cross sections for $b\bar{b}$ annihilation and $gg$
fusion through a top loop, $gg\toploop h(A),$ in the SM and MSSM. The $b%
\bar{b}\rightarrow h(A)$ process is calculated as described above. The
process $gg\toploop h(A)$ is calculated using the HIGLU program~\cite%
{Spira:1996if}, which includes the diagrams of Fig.~\ref{fig:bbh-ggh}. %
\figbbhggh

The cross sections for the Tevatron and LHC as a function of the Higgs mass
are shown in Fig.~\ref{HiggTev}\figHiggTev and Fig.~\ref{HiggLhc},
respectively. In
these figures, the dashed lines represent the cross sections for the $b\bar{b}%
\rightarrow H(A)$ process; the dotted lines the $gg\toploop H(A)$ process;
and the solid lines the combined results. Both renormalization and
factorization scales are set to be $%
M_{H(A)}$, and the PDFs used are the CTEQ6M set~\cite{Pumplin:2002vw}. For
the MSSM case, the results correspond to $\tan \beta =10$ and $m_{%
\mathrm{top}}=178$~GeV. \figHiggLhc

One can see that in the SM case, the contribution from \bbH is negligible
compared to \ggH. In contrast, the contribution from the Supersymmetric \bbA
process becomes important even for the moderate value
of $\tan \beta \sim 10$. Except for the low Higgs mass region $M_{H}<115$%
~GeV, the $b\bar{b}\rightarrow A$ process is the dominant production
mechanism. The ratio of \bbA to \ggA processes is qualitatively similar at
higher values of $\tan \beta$ while the absolute value of the cross sections
scales as $\tan^2\beta$. Relative ratios of the $b\bar{b}\rightarrow A$ and %
\ggA processes at the Tevatron and LHC are very similar, while the absolute
values of the production rate at the LHC is about two orders of magnitude
higher then those at the Tevatron. The cross section for both processes is
enhanced with high values of $\tan \beta $ and can be really large.
Therefore, these processes could be useful for precision measurement
of Yukawa couplings as well as bottom-quark distributions. This underlines
the importance of understanding the PDF uncertainties of the cross sections
for these processes.



\section{Uncertainties of the Cross Sections for Higgs Production}
\label{sec:Uncert}

 The PDF uncertainties for Higgs production are most
reliably assessed by the Lagrange multiplier (LM) method
\cite{Pumplin:2000vx,Stump:2001gu}. It incorporates the calculated values of
the Higgs cross section $\sigma$ in the global analysis, using the classic
Lagrange multiplier technique. This determines the full allowed range of
variation of $\sigma$ over the PDF parameter space. Our main results are
obtained by this method. An alternative approach that can be applied without
performing dedicated global analysis is the Hessian method
\cite{Pumplin:2000vx,Pumplin:2001ct}, which
has been applied to the Higgs production \ggH process in~\cite%
{Djouadi:2003jg} and to $gg\to H\bar{b} b$ in~\cite{chris-pdf}. The Hessian
method is less robust because it relies on a linear approximation in the
error analysis. As a part of our study, we will compare results obtained by
the two methods. A brief summary of both methods is given in the Appendix.

\subsection{PDF Uncertainties for the \bbA process: the Lagrange Multiplier
Method}

For a given Higgs mass, the result from a LM study of the range of variation
of the Higgs cross section is presented in Fig.~\ref{fig:LmScan}. The plot
shows the goodness-of-fit of the global analysis (as measured by an overall
effective $\chi ^{2}$ value) in constrained fits, as a function of the Higgs
cross section (for the \bbA process) over a certain range around
the best-fit value. The curves represent smooth interpolations of these
constrained fits (cf.\ the Appendix, around
Eqs.~(\ref{eq:lmf}-\ref{eq:dx}),
for explanation of the method). Fig.~\ref{fig:LmScan}(left) presents results
for Tevatron for $M_{A}$=100~GeV, and Fig.~\ref{fig:LmScan}(right) presents
results for LHC for $M_{A}$=400~GeV. These curves are quite close to being
parabolic. This suggests that the alternative Hessian method may be a
reasonable approximation.\figLmScan

The uncertainty range of the Higgs cross section $\sigma $ is obtained by
adopting a reasonable \emph{tolerance} for the global $\chi ^{2}$,
$T^2=\Delta \chi ^{2}$. \ Various global analysis groups (CTEQ, MRST, ZEUS,
H1) have adopted values of $T^2$ in the range $50$-$100$, for $\sim 2000$
data points. We shall choose the more conservative value $T^2=100$, which we
interpret as a 90\% CL uncertainty range \cite{Pumplin:2002vw}. We obtain
$\sigma _{+}$ and $\sigma _{-}$ as the two solutions of the equation $\chi
^{2}[\sigma ]=T$; and define the PDF uncertainty of the total cross section
as
\begin{equation}
\Delta \sigma =(\sigma _{+}-\sigma _{-})/2,
\end{equation}%
and the relative uncertainty as
\begin{equation}
\delta\sigma=\Delta \sigma/\sigma \,.
\end{equation}%

We evaluated this relative uncertainty of the Higgs cross section, due to the
input PDFs, for various values of the Higgs mass.\figLmResults
 The results are
presented in Fig.~\ref{fig:LmResults}(left) for the Tevatron, and Fig.~\ref%
{fig:LmResults}(right) for the LHC, as solid lines. For comparison, QCD scale
uncertainties available for the Higgs mass range below 300~GeV from
Ref.~\cite{Harlander:2003ai} are represented by dashed lines.

As one can see, there is a qualitative difference in the behavior of ${\delta
\sigma ^{PDF}}$ as a function of Higgs mass between the Tevatron and LHC
results: at the Tevatron, ${%
\delta \sigma ^{PDF}}$ always increases with increasing Higgs mass; while at
the LHC, it has a minimum for Higgs mass around 300 GeV. To understand the
reason for this behavior one can look at the uncertainty of the $gg$
luminosity function, which is directly related to the $b$-quark PDF
uncertainty, since gluon splitting creates the $b$-quark parton density. This
is shown in Fig.~\ref{fig:Glum}, for the Tevatron on the left and LHC on the
right as a function the gluon-gluon invariant mass. \figGlum One can see that
for $M_{A}>100$~GeV, the uncertainty of $gg$ luminosity always goes up with
\figPdfUnX the increasing Higgs boson mass, which is related to the fact that
at the Tevatron the required $x$-value of PDF is \textit{already} as big as
0.05 ($x=M_{A}/\sqrt{S}\gtrsim 100/1960$). Therefore with the increasing
Higgs boson mass, $x$ goes up and so does ${\delta \sigma ^{PDF}}$. At the
LHC,
however, for low $M_{A}\simeq 100$~GeV, $x_{min}\simeq 0.007$ and therefore $%
\delta \sigma ^{PDF}$ is \textit{still} big since $x$ is fairly small. When
the Higgs mass increases and reaches $M_{A}\simeq 300$~GeV, ${\delta \sigma
^{PDF}}$ takes the minimum at $x_{min}\simeq 0.02$. With further Higgs mass
increase, ${\delta \sigma ^{PDF}}$ grows similarly to its behavior at the
Tevatron.

Actually, the $x$-value is the principal variable that controls the PDF
uncertainty. This can be clearly seen from Fig.~\ref{fig:PdfUnX} which
presents ${\delta \sigma ^{PDF}}$ as a function of $x$ for Tevatron and LHC:
in the $x$-region where Tevatron and LHC overlap, their PDF uncertainties are
in good agreement.

\subsection{Comparison of PDF to Scale Uncertainties}

Included in Fig.~\ref{fig:LmResults} is a direct comparison of PDF
uncertainties ${\delta \sigma ^{PDF}}$ of the Higgs cross section with the
scale uncertainty ${\delta \sigma ^{SC}}$. The latter has been obtained from
the QCD scale dependence of the next-to-next-leading order (NNLO)
calculation of the inclusive \bbA$%
+X$ processes~\cite{Harlander:2003ai}. It was found that the scale
uncertainty goes down from $15\%$ to $5\%$ at the LHC and from $10\%$ to $%
3\% $ at the Tevatron when the Higgs mass increases from $120$ to $300$~GeV.
We notice the opposite trend of those uncertainties versus Higgs boson mass
at the Tevatron: ${\delta \sigma ^{PDF}}$ goes \textit{up} from 11\% to
about 30\% for $M_{A}$ increasing from 100 to 200~GeV, while ${\delta \sigma
^{SC}} $ \textit{decreases} from $11\%$ to $6\%$. Therefore, at high $M_{A}$
values, ${\delta \sigma ^{PDF}}$ becomes almost an
order of magnitude larger than $%
{\delta \sigma ^{SC}}$!

At the LHC, both ${\delta \sigma ^{PDF}}$ and ${\delta \sigma ^{SC}}$
decrease with the increasing $M_{A}$ in this mass range ($100-200$~GeV);
but ${%
\delta \sigma ^{SC}}$ is larger than ${\delta \sigma ^{PDF}}$ by a factor $%
1.5-3$, depending on the Higgs mass. This plot suggests that at higher values
of $M_{A}$, say $>300$ GeV, the PDF uncertainties will become dominant,
similar to the situation at the Tevatron. However, NNLO scale
uncertainties were not published for this Higgs mass range.\figBbUnHs

\subsection{Comparison of the Lagrange Multiplier and Hessian Methods for
Estimating Uncertainties}

We calculated the PDF uncertainties of the cross section using the LM
method, which is the most reliable one available. But it requires the full
machinery of global analysis. The alternative Hessian method, utilizing a
general set of eigenvector PDF sets that embody the PDF uncertainties
\cite{Pumplin:2002vw}, is more approximate, but more convenient
\cite{Pumplin:2000vx,Pumplin:2001ct}.
Fig.~\ref{fig:BbUnHs} presents the results
on PDF uncertainties of the \bbA cross section obtained by the Hessian
method, compared to that obtained by the LM method for Tevatron (left) and
LHC (right). As one can see, the two results are in good agreement. In
Fig.~\ref{fig:GgUnHs} we present results analogous to Fig.~\ref{fig:BbUnHs}
but for the \ggA process. Again, one can see that there is good agreement
between the two methods.\figGgUnHs

\subsection{Comparison of PDF Uncertainties of the \bbA and \ggA Processes}

Notice that PDF uncertainties for the \ggA process in comparison with \bbA
are about a factor of two smaller. To understand this fact we refer the
reader to~\cite{Sullivan:2001ry} where $\delta g/g$ and $\delta b/b$
uncertainties and their correlation were studied in detail. For the
Tevatron, PDF uncertainties vary from $5$ to $15-20\%$ for $M_{A}$
ranging
between $100$ and $200$~GeV, and dominate the scale uncertainty only for
heavy Higgs of mass about $150-160$~GeV. For the LHC, PDF uncertainties are $%
5-6\%$ at $M_{A}=100$~GeV, decreasing to the minimum of $3-4\%$ at $M_{A}=300
$~GeV and increasing again up to about 11\% at $M_{A}=1000$~GeV. Available
scale uncertainties for $M_{A}<300$~GeV are about a factor of two bigger
than PDF uncertainties. Our results on \ggA PDF uncertainties are in
agreement with results presented in Refs.~\cite{Djouadi:2003jg,chris-pdf}.

\figCorGgBb We note that the PDF uncertainties for \bbA and for \ggA are
strongly correlated, as shown in Fig.~\ref{fig:CorGgBb}, both for the
Tevatron and for the LHC. This is hardly surprising, given the fact that the
$b$-quarks are radiatively generated from the gluon in the way all current
parton distribution functions are calculated.

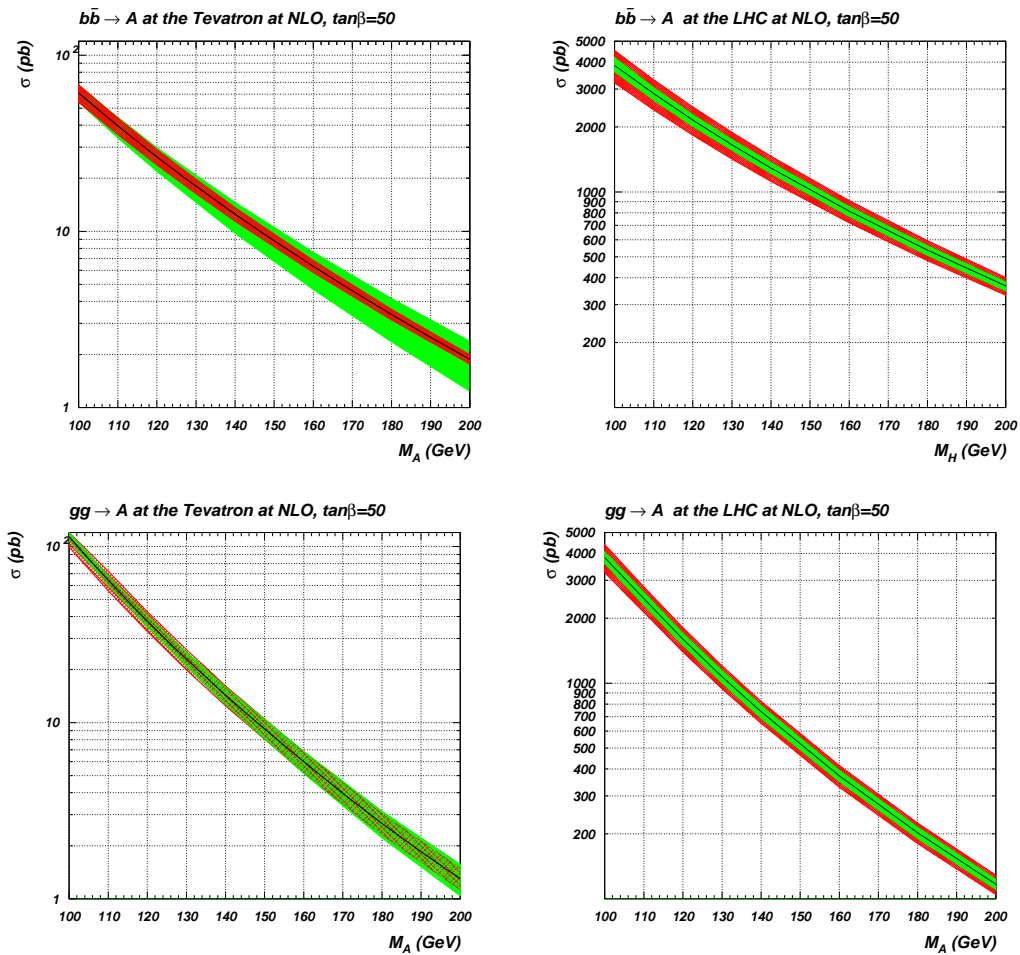
Finally, Fig.~\ref{fig:final-cs} illustrates the NLO cross section bands for
PDF uncertainties, shown in green, overlaid with the NNLO scale
uncertainties, shown in red. One can see that for low Higgs mass, the scale
uncertainties are comparable or dominant for both colliders and both
processes, while for heavier Higgses, PDF uncertainties are significantly
larger than the scale uncertainties.

\figFinalCs

\vspace*{-0.3cm}
\section{Conclusions}
\label{sec:Concl}
\vspace*{-0.2cm}

The role of \bbA and \ggA processes
may be  central for the Higgs boson search.
Therefore the correct understanding of uncertainties of their production rate
 is crucial.

We found that the PDF uncertainty of \bbA is about a factor two larger than
the PDF uncertainty for \ggA. It was found that at the Tevatron, PDF
uncertainty dominates the scale uncertainty for $M_A>$130~GeV and could be as
large as $30\%$ for $M_A$=200~GeV, which is an order of magnitude larger than
the NNLO scale uncertainty. At the LHC the scale uncertainty is dominant and
could be as big as 15\% for $M_A<$300~GeV. In this region one could expect
large Higgs production rates that would statistically allow  the precision
measurement of Higgs Yukawa couplings. Therefore, higher order corrections
would be necessary in this case for better theoretical control of the cross
section. For $M_H>$300~GeV, PDF uncertainty is likely to dominate at the LHC,
similarly to the picture for the Tevatron. These results underline the
importance of gaining better control of the PDF uncertainties, in the study
of Higgs physics in the next generation of Colliders.

We have also found that the Lagrange Multiplier and Hessian methods for
assessing PDF uncertainties are in good agreement with each other.

\section*{Acknowledgments}
We thank M.~Spira  for useful discussions. This work was supported in part by
the U.S. National Science Foundation under awards PHY-0354838 and PHY-0244919.
C.-P.Y. thanks the hospitality of National Center for Theoretical Sciences in
Taiwan, R.O.C., where part of this work was performed.


\section*{APPENDIX}

In this appendix we briefly review the methods of global PDF fit and
estimation of PDF uncertainties using Hessian and Lagrange multiplier
methods.

Parton distributions which are being used for the SM and new physics
predictions are obtained from global analysis using a ``best-fit'' paradigm for
which the PDF is selected for the minimum of the chosen $\chi^2$ function.
The main question is what are the uncertainties of those PDFs?

In our study we are using two methods for this purpose, namely Hessian
method~\cite{Pumplin:2000vx,Pumplin:2001ct} and method of Lagrange
multiplier~\cite{Stump:2001gu} which as it was discussed in Ref.~\cite%
{Pumplin:2002vw}, overcomes various problems of standard error analysis. In
particular, in Ref.~\cite{Pumplin:2002vw}, the authors presented a 
reliable way
of understanding the behavior of $\chi^2$ function in the neighborhood
of the global minimum, providing the way of correct understanding of the PDF
uncertainties in the prediction of the cross sections.

We summarize here briefly both methods. Both methods use a chi-square
function $\chi^{{2}}$ is defined by
\begin{eqnarray}
\chi^2 &=& \sum_{e} \chi_{e}^{2}(a,r), \mbox{\ \ \ where \ \ \ }
\chi_{e}^{2}(a,r) = \sum_{i} \frac{\left[D_{i}-\sum_{k}r_{k}\beta_{ki} -
T_{i}(a)\right]^{2}} {\alpha_{i}^{2}} +\sum_{k}r_{k}^{2},
\end{eqnarray}
where $e$ labels an experimental data set and $i$ labels a data point in each
particular data set. $D_{i}$ is the data value, $\alpha_{i}$ is the
uncorrelated error, and $\beta_{ki}$ is the $k$th correlated systematic
error; these numbers are published by the experimental collaboration. $%
T_{i}(a)$ is the theoretical value, a function of a set of $n$ PDF
parameters, $\{a_{1}, \dots , a_{n}\}$. Also, $\{r_{k}\}$ is a set of
Gaussian random variables and $r_{k}\beta_{ki}$ is a (correlated) shift
applied to $D_{i}$ to represent the $k$th systematic error. We minimize the
function $\chi^{{2}}(a,r)$ with respect to both the PDF parameters
$\{a\}$ and the systematic shift variables $\{r_{k}\}$. The result yields
both the standard PDF model with parameters $\{a_{0}\}$, and the optimal
shifts $\{\widehat{r}_{k}\}$ to bring theory and data into agreement. This
minimum of $\chi^{{2}}$ represents the best fit to the data~\cite%
{Pumplin:2002vw}.

Hessian method for analysis of PDF uncertainty in the neighborhood of the
minima of $\chi^2$ involves the Hessian matrix
\begin{equation}
H_{ij}=\frac{1}{2}\frac{\partial^{2}{\chi}^{2}_0} {\partial{a}_{i}\partial{a}%
_{j}}
\end{equation}
calculated at the minimum of $\chi^2_0$. The next step is to diagonalize $%
H_{ij}$ and to find its eigenvectors. Then for each eigenvector we
have two displacements from $\{a_{0}\}$ (in the $+$ and $-$ directions along
the vector) denoted $\{a^{+}_{i}\}$ and $\{a^{-}_{i}\}$ for the $i$th
eigenvector. At these points, ${\chi}^{2}_{\pm}={\chi}_{0}^{2}+T^{2}$ and $T$
parametrizes the \emph{tolerance}. The appropriate choice of tolerance $T$
cannot be decided without a further, more detailed, analysis of the quality
of the global fits. After studying a number of examples~\cite%
{Pumplin:2001ct,Stump:2001gu}, we concluded that a rather large tolerance, $T
\sim 10$, represents a realistic estimate of the PDF uncertainty.

One can show that in a linear approximation, the uncertainty $\delta X$ for
any quantity $X$, which depends on PDF, can be expressed as
\begin{equation}
\left(\delta{X}\right)^{2} = T^{2} \sum_{i,j} \left(H^{-1}\right)_{ij} \frac{%
\partial{X}}{\partial{a}_{i}} \frac{\partial{X}}{\partial{a}_{j}};
\end{equation}
or, in terms of the eigenvector basis sets,
\begin{equation}  \label{eq:ME1}
\left(\delta{X}\right)^{2} = \frac{1}{4}\sum_{k=1}^{n} \left[%
X(a^{+}_{i})-X(a^{-}_{i})\right]^{2}.
\end{equation}
representing the master equation defined in~\cite{Stump:2001gu}. One should
point out again, that equation (\ref{eq:ME1}) is based on a linear
approximation: ${\chi}^{2}(a)$ is assumed to be a quadratic function of the
parameters $\{a\}$, and $X(a)$ is assumed to be linear. This approximation
is not strictly valid in general.

The essence of the Lagrange multiplier method is the introduction of the
Lagrange multiplier variable $\lambda$ and minimizing the function
\begin{equation}
\chi^2_\lambda (\lambda ,a)=\chi^{2}(a)+\lambda X(a)  \label{eq:lmf}
\end{equation}
with respect to the original $n$ parameters $\{a\}$ for fixed values of $%
\lambda$. In Eq.\ref{eq:lmf} X(a) is some observable as in the example for
Hessian method. Minimization of $\chi^2_\lambda (\lambda ,a)$ for various
values of $\lambda$ allows to find the parametric relationship between $%
\chi^{2}(a)$ and $X(a)$, .i.e.
\begin{equation}
\chi^2_\lambda (\lambda ,a_0)=\chi^{2}(a_0)+\lambda X(a_0) \Longrightarrow
X=X(\chi^{2}(a_0,\lambda)),  \label{eq:lmf1}
\end{equation}
where $a_0$ is the set of parameter values $\{a\}$ for each particular value
of $\lambda$. Eq.~(\ref{eq:lmf1}) is the key point of LM method since for
given value of tolerance
\begin{equation}
\Delta\chi^2=\chi^{2}(a_0,\lambda^\Delta_\pm)-\chi^{2}(a_0,0)
\end{equation}
which would correspond to some two values $\lambda^\Delta_\pm$, one can find
the respective variation of the observable $X$:
\begin{equation}
\delta X_+=X(\chi^{2}(a_0,\lambda^\Delta_+))-X(\chi^{2}(a_0,0)), \ \ \delta
X_-=X(\chi^{2}(a_0,\lambda^\Delta_-))-X(\chi^{2}(a_0,0)) \,. \label{eq:dx}
\end{equation}
The  LM method for calculating $\delta X_\pm$ is more robust in general
since it does not approximate $X(a)$ and $\chi^{2}(a)$ by linear and
quadratic dependence on $\{a\}$, respectively, around the minimum.


\bibliographystyle{JHEP}
\newpage
\bibliography{bbh}

\end{document}